\documentstyle[pre,aps,float,eqsecnum,epsf,multicol]{revtex}

\newcommand{\al}{\alpha}
\newcommand{\be}{\beta}
\newcommand{\fbar}{\bar{f}}

\newcommand{\ts}{T^{\ast}}
\newcommand{\rs}{\rho^{\ast}}

\newcommand{\dee}{\partial}
\newcommand{\xt}{\kappa a_2}

\newcommand{\app}{\rightarrow}
\newcommand{\xtee}{x_{\mbox{\scriptsize T}}}
\newcommand{\xeks}{x_{\mbox{\scriptsize X}}}
\newcommand{\xoh}{x_{\mbox{\scriptsize O}}}

\newcommand{\kbj}{K^{Bj}}
\newcommand{\kbjt}{K^{Bj}(T)}
\newcommand{\keb}{K^{Eb}}
\newcommand{\kebt}{K^{Eb}(T)}
\newcommand{\half}{\frac{1}{2}}
\newcommand{\etwo}{\varepsilon_2}
\newcommand{\cvcon}{C_V^{\mbox{\scriptsize conf}}}

\begin{document}
\title{Critique of Primitive Model Electrolyte Theories}
\author{Daniel M. Zuckerman, Michael E. Fisher, and Benjamin P.
Lee\footnote{Current address: Polymers Division, National Institute of
Standards and Technology, Gaithersburg, MD 20899}}
\address{Institute for
Physical Science and Technology, University of Maryland, College Park,
Maryland 20742}
\date{May 8, 1997}
\maketitle

\begin{abstract}
\indent Approximate theories for the restricted primitive model electrolyte are
compared in the light of Totsuji's lower bound for the energy (an improvement
over Onsager's), Gillan's upper bound for the free energy, and thermal
stability requirements.
Theories based on the Debye-H\"{u}ckel (DH) approach and the mean spherical
approximation (MSA), including extensions due to Bjerrum, Ebeling, Fisher and
Levin, and Stell, Zhou, and Yeh (PMSA1, 2, 3) are tested.
In the range $\ts = k_B T D a/q^2 \lesssim 10 \, \ts_c \simeq 0.5$, all
DH-based theories satisfy Totsuji's bound, while the MSA possesses a
significant region of violation.
Both DH and MSA theories violate Gillan's bound in the critical region and
below unless ion pairing {\em and } the consequent free-ion depletion are
incorporated.
However, the PMSA theories, which recognize pairing but not depletion, fail to
meet the bound.
The inclusion of excluded-volume terms has only small effects in this respect.
Finally, all the pairing theories exhibit negative constant-volume specific
heats when $\ts \gtrsim 2 \, \ts_c \simeq 0.1$;
this is attributable to the treatment of the association constant.
\end{abstract}

\bigskip

\begin{multicols}{2}

\section{Introduction}
The liquid-gas phase transition in electrolytes is of current interest because
of puzzling experiments and theoretical efforts to understand them.
For recent reviews, see \cite{levelts,fisher,stellrev}.
The primary model used is the restricted primitive model (RPM) consisting of
two oppositely charged, but otherwise identical, sets of $N_+ = N_-$ hard
spheres of diameter $a$ and charge per particle $\pm q$, immersed in a medium
of dielectric constant $D$ (to represent the solvent) and volume $V$.
We will restrict our attention to the RPM in $d=3$ dimensions and use the
reduced temperature and density
\begin{equation}
T^{\ast} = k_B T D a/q^2 \;\;\;\;\; \mbox{and}
	\;\;\;\;\; \rho^{\ast} = a^3 \rho,
\end{equation}
where $\rho = (N_+ + N_-)/V \equiv N/V$; the Debye inverse screening length,
\begin{equation}
\label{eqkapx}
\kappa_D = ( 4 \pi q^2 \rho /D k_B T )^{1/2}, \;\; \mbox{with} \;\;\;
	x = \kappa_D a = \surd(4 \pi \rs / \ts) \, ;
\end{equation}
the reduced Helmholtz free energy density
\begin{equation}
\fbar(\rho,T) = - F_N(V;T)/V k_B T;
\end{equation}
and the reduced configurational energy per particle, $u$, defined via
\begin{equation}
\label{equdef}
(Nq^2/Da)\,u(\rho,T) = U_N - \hbox{$\frac{3}{2}$} N k_B T,
\end{equation}
where $F_N$ and $U_N = V k_B T^2 \, (\partial \fbar / \partial T)$ denote the
total free energy and (internal) energy.

Recent theory\cite{fisher,stellrev,fl,lf,gandg,evans,stellassoc} has focussed
on two approaches to approximating the free energy of the RPM, based on either
Debye-H\"{u}ckel (DH) theory\cite{debye} or the mean spherical approximation
(MSA)\cite{msa,wais,gmsa}.
Many years ago Bjerrum\cite{bj} proposed to improve DH theory by including ion
pairing via ``chemical association.''
Later, Ebeling and Grigo\cite{ebgr} combined ion-pairing with an MSA expression
for the ionic free energy; more recently, Levin and Fisher \cite{lf} and Stell
and coworkers\cite{stellassoc} explored further extensions of the MSA.
On the other hand, Fisher and Levin\cite{fl,lf} supplemented DH theory not only
with ion pairing and excluded-volume terms but also included the solvation free
energy of the electrically active $(+,-)$ dipolar ion pairs.
Currently, this class of DH-based theories seems to give the best, albeit
semiquantitative, account of the RPM in the critical region as judged by
comparison with simulations performed by various authors\cite{fisher,lf}.
It may be remarked that the simulation estimates for $\ts_c$ and $\rs_c$ have
been changing at an alarming rate [2(b)].
Nevertheless, the MSA-based theories yield approximations for
$\ts_c \, ( \, \gtrsim 0.073)$
that are significantly {\em higher} than those based on the DH approach
($\ts_c \lesssim 0.056$),
which in fact agrees much better with the simulations ($\ts_c = 0.048 \mbox{-}
0.056$)\cite{fisher,lf,stellassoc,orkpan}.

At a purely theoretical level, however, one cannot be content since, {\em a
priori}, there seem no clear grounds for preferring the DH-based theories ---
apart from their more direct and intuitive physical interpretation --- rather
than the more modern (and fashionable) MSA-based theories which --- since they
entail the pair correlation functions and the Ornstein-Zernike (OZ) relation
--- give the impression of being more firmly rooted in statistical mechanics.
On the other hand, it has recently been shown that the DH theories yield pair
correlations satisfying the OZ relation in a very natural way\cite{benf}.
Furthermore, both theories have an essentially mean-field character despite
which, in contrast to typical mean field theories for lattice systems, {\em
neither} has any known Gibbs-Bogoliubov variational formulation or similar
basis.
How, then, might the two approaches be distinguished?

Now Blum and his coworkers have, in various
places\cite{blum,bernard,blumnew,simblum}, enthusiastically sung the praises of
the MSA for the RPM, asserting that the theory ``$\,$is asymptotically correct
in the limit of high density and infinite charge'' or ``high screening
parameter (Debye length going to zero).''
Furthermore, ``unlike the DH theory, it [the MSA] satisfies the exact Onsager
bounds for the Helmholtz free energy and the internal
energy''\cite{blum,bernard} (in the same asymptotic limit) and the ``internal
energy of the MSA is an exact lower bound''\cite{blumnew}.
As reported below, these claims cannot be sustained:
however, they do suggest that one might usefully assess and compare the MSA and
DH theories, and their various extensions, by checking their predictions
against previously developed bounds for the internal energy and Helmholtz free
energy.
That task is undertaken here.

Indeed, as discussed more fully in Sec. $\!\!$II, several bounds have been
established.
The well-known Onsager {\em lower} bound for the {\em configurational energy}
of the RPM was derived in 1939 \cite{onsag};
less heralded is an improvement due to Totsuji some forty years later
\cite{tot}.
For the {\em free energy}, Rasaiah and Stell \cite{stell} proved that the
hard-core free energy provides an {\em upper} bound, while Gillan \cite{gil}
developed a much stronger upper bound embodying the idea of $(+,-)$ pairing
into dipoles\cite{fisher,fl,lf,bj,ebgr}.
Finally, we note that thermodynamic stability with respect to temperature
requires the positivity of the specific heat at constant volume \cite{pippard}.

We will focus particularly on the Totsuji and Gillan bounds applied in the
region of the predicted gas-liquid phase transition and critical point.
We find that DH theory and all its augmentations always {\em satisfy} Totsuji's
(and Onsager's) bound provided $\ts \lesssim 10 \ts_c \simeq 0.5$.
On the other hand, the MSA actually {\em violates} the Totsuji bound in a
significant region of the $(\rho,T)$ plane where coexistence is predicted,
unless the theory is suitably augmented.

In the light cast by Gillan's bound, the two approaches rest on a more equal
footing.
As already shown by Gillan \cite{gil}, the MSA (in its usual form) fails badly
for
$\ts \lesssim 0.08$;
but the same is true for the original DH theory (even when supplemented by
excluded-volume terms \cite{fisher,fl,lf}).
Only when both basic theories are augmented by ion-pairing contributions and by
allowing for the associated depletion of the free-ion screening do they satisfy
the Gillan bound.
As against the hard-core electrostatic effects, included in {\em both} DH and
MSA treatments, the presence or absence of specific excluded-volume terms has
small effect numerically and does not affect the satisfaction of the bound.
However, the recent PMSA (or pairing-MSA) theories of Stell and coworkers
\cite{stellassoc} violate Gillan's bound apparently because they do not account
appropriately for the free-ion depletion.

The main lesson is the crucial importance of the clustering of ions into
dipolar pairs at low temperatures.
Of course, this has been appreciated heuristically for a long time \cite{bj}
and was quantitatively demonstrated in 1983 by Gillan \cite{gilsim} in
calculations for the RPM which showed that the vapor for
$\ts \lesssim 0.053$
consisted mainly of $(+,-)$, $(2+,2-)$, $(3+,3-)$, $\ldots$ neutral clusters
and $(2+,1-)$, $(1+,2-)$, $(3+,2-)$, and $(2+,3-)$ singly charged clusters,
with relatively far fewer free monopoles, $(+)$ and $(-)$.
The present work, however, seems to be the first purely analytic demonstration
of the thermodynamic necessity for including clustering, implicitly or,
perhaps, explicitly, in approximate theories.

The recognition of $(+,-)$ ion-pairing requires the specification of the
corresponding {\em association constant}, $K(T)$.
Ever since Bjerrum's original proposal \cite{bj}, this has been a matter of
confusion and contention (see, e.g., \cite{fisher,fl}).
Nevertheless, in the low temperature region of principal interest here, say
$\ts \lesssim 0.08 \simeq 1.5 \, \ts_c$, Bjerrum's cutoff form and Ebeling's
more sophisticated expression agree to within 1.8\% or better
\cite{fl,lf,bj,ebgr} and, along with other cutoff forms, have identical
asymptotic expansions in powers of $\ts$ \cite{lf,fuoss}.
For practical purposes, therefore, $K(T)$ might be regarded as known
``exactly.''
At higher temperatures, where pairing should be (and is predicted to be) much
weaker, it is natural to surmise that different treatments of association would
prove inconsequential.
However, this proves false!
Indeed, for all the previous pairing theories
\cite{fl,lf,evans,stellassoc,bj,ebgr} we find that the constant-volume
configurational specific heat becomes {\em negative} (violating thermodynamics
\cite{pippard} and statistical mechanics) in the region $\ts = 0.1 \; \mbox{to}
\; 0.5$: see Sec. IV.
The source of this serious problem is found in the proposed behavior of the
association constant.
Initial steps towards amelioration are indicated, but the issue will be pursued
in more detail elsewhere \cite{assocpaper}.

It should be mentioned that we also examine the generalized MSA (GMSA)
\cite{evans,gmsa} and variants of the MSA thermodynamics derived from the
(approximate) pair correlation functions by routes other than the standard
energy equation \cite{wais}:
these are discussed in Sec. III.
Other even less realistic models for electrolytes exist, including the
one-component plasma with hard cores \cite{palmer} and the corresponding
``dense-point limit'' [11(c)];
however, we address here only the RPM.

The explicit comparisons of the DH and MSA theories {\em without} allowance for
ion pairing are presented in Sec. III, below.
In Sec. IV the theories that include descriptions of ion pairing are assessed,
including the PMSA theories \cite{stellassoc}.

\vfil\eject

\section{Bounds for the Energy and Free Energy}

\subsection{Configurational Energy Bounds}
The first rigorous lower bound for the configurational energy of the RPM seems
to be due to Onsager \cite{onsag}.
It is essentially a consequence of the positivity of the total electrostatic
potential energy density and, with the notation of (\ref{equdef}), yields
\begin{equation}
\label{eqonsag}
u(\rho,T) \geq u_{\mbox{\scriptsize Ons \normalsize}} = -1.
\end{equation}
A more transparent derivation for a system with a neutralizing background has
been presented by Rosenfeld and Gelbart \cite{onsagdemo}.
Totsuji, in 1981 \cite{tot}, improved on Onsager's result by writing the energy
as an integral over the ionic pair correlation functions and showing that the
presence of the hard-core repulsions implies an upper bound on the correlation
functions.
He thence established
\begin{equation}
\label{eqtot}
u(\rho,T) \geq u_{\mbox{\scriptsize Tot \normalsize}} = -0.960.
\end{equation}
Although the improvement is by only $4.0 \%$, it has significant consequences.

As remarked by Totsuji, one may usefully compare these bounds with the
electrostatic or Madelung energies of an ionic crystal;
for the NaCl (sc) and CsCl (bcc) structures one has \cite{kittel}
\begin{equation}
\label{eqxtal}
u_{\mbox{\scriptsize NaCl \normalsize}} \simeq -0.8738 \; \; \; \; \;
\mbox{and} \;\;\;\;\; u_{\mbox{\scriptsize CsCl \normalsize}} \simeq -0.8813 .
\end{equation}
One may reasonably suppose that the latter represents the best possible lower
bound and so we will also invoke it in testing approximate theories for the
RPM.

\subsection{Gillan's Free Energy Upper Bound}
Gillan \cite{gil} has developed a convincing, but not fully rigorous, upper
bound on the Helmholtz free energy of the RPM, which incorporates the idea of
ion pairing.
The pure hard-core free energy actually provides a rigorous upper bound
\cite{stell}, but Gillan's bound is lower except for extremely low densities
$(\rs \lesssim 10^{-5})$ where the limiting behavior is well understood.
Here we utilize only Gillan's bound, which is derived with the aid of the
Gibbs-Bogoliubov inequality by employing a sequence of truncated reference
systems.
The calculation finally incorporates paired $(+,-)$ ions or dipoles by using a
reference system of over-sized, spherically-capped cylinders with modified
Coulomb interactions.
The last step of Gillan's argument relies on a comparison of an approximate
analytical expression for the pressure of a system of such spherocylinders with
computer simulation estimates\cite{nezbeda,rebertus}:
the approximate formula appears to provide a bound on the true results.
A search of the more recent literature concerning this system (e.g. Refs.
\cite{citer1,citer2,levesque,citer3}) indicates that the original simulations
have withstood the test of time.
(However, Frenkel \cite{frenkel} has observed that at high densities and for
length/diameter ratios larger than needed here, the simulations --- and,
certainly,  the analytic approximation --- miss an isotropic-nematic fluid
transition that is to be expected.)
We thus believe that Gillan's bound is valid.

To display the bound explicitly, we write the diameter and the chosen
\cite{gil} center-to-center distance of the spherocylinders as $a_s = (1 +
\delta) a$ and put
\begin{equation}
\lambda \equiv (5 \pi/24) \rho a_s^3 = (5 \pi/24)(1 + \delta)^3 \rs.
\end{equation}
If $\fbar^{\, \mbox{\scriptsize Id \normalsize}} \!\! (\rho,T)$ is the
ideal-gas free energy density, we then have \cite{gil}
\begin{equation}
-\fbar(\rho,T) \leq -\fbar^{\, \mbox{\scriptsize Id \normalsize}} \!\! (\rho,T)
+ \hbox{$\frac{1}{2}$} \, \rho \, {\cal F}(\rho,T),
\end{equation}
\begin{equation}
{\cal F}(\rho,T) = 1 - 2 \pi \rs - \frac{1}{\ts}
   - \hbox{$\frac{18}{5}$} \, \lambda \, \frac{1 - \frac{2}{5} \lambda}{(1 -
\lambda)^2}
   -\ln{{\cal L}(\rho,T)},
\end{equation}
\begin{equation}
{\cal L}(\rho,T) = \ts (1 - \lambda) \left \{ 1 -
	\exp{[-\delta/\ts(1 + \delta)]} \right \} .
\end{equation}
We will adopt $\delta = 0.3$, which Gillan found optimized the bound for most
values of $T$.

\section{Basic Theories for the RPM: Comparison with Bounds}
\subsection{DH and MSA without pairing}
\subsubsection{DH theory}
Debye-H\"{u}ckel theory\cite{debye} (here referred to as ``pure'' DH theory,
since explicit dipolar pairing is not included) is the oldest theory for
electrolytes still in current use.
The theory entails two approximations:
first, the pair correlation functions, $g_{ij}({\bf r}_i - {\bf r}_j)$, are
represented by naive Boltzmann factors --- with the charge, $q_j$, multiplied
by the average electrostatic potential at ${\bf r}_j$ when an ion of charge
$q_i$ is fixed at ${\bf r}_i$ --- ignoring higher order correlation effects;
and, second, these Boltzmann factors are linearized, which is valid only in the
limit of low density, small charge, or high temperature.
(For a modern discussion, see McQuarrie \cite{debye}.)
The thermodynamics predicted by DH theory depends only on the single parameter,
$x= \kappa_D a$.
The appearance of the hard core diameter, $a$, demonstrates that DH theory
takes account of the electrostatic effects of the hard cores;
however, the original or pure DH theory did not treat the excluded-volume
effects of the hard cores (and so reduced to a theory for an ideal gas mixture
in the limit of vanishing charge, $q \rightarrow 0$).
Nonetheless, excluded volume contributions may be included naturally by adding
to the free energy a suitably chosen pure hard-core term \cite{fl,lf}; see
below.
In the DH critical region, such terms have a relatively small effect.

\subsubsection{The MSA and variants}
The other ``basic'' theory we consider, the mean spherical
approximation\cite{msa}, is defined by a closure of the Ornstein-Zernike
relation in which the $g_{ij}({\bf r})$ vanish {\em inside} the hard core,
while the direct correlation functions {\em outside} the hard-core exclusion
zone are approximated by the Coulombic potentials.
Waisman and Lebowitz\cite{wais} solved the MSA exactly for the RPM: that is,
they determined the correlation functions which, in principle, yield the
thermodynamics.
The electrostatic free energy again depends only on $x = \kappa_D a$, but it
and the overall free energy depend strongly on the theoretical route taken ---
via, in particular, the energy, pressure, or compressibility relations.
Since very different results are obtained, we review them briefly.
The standard MSA thermodynamics almost invariably discussed in the literature
employs the energy route; but as a result, {\em no excluded-volume hard-core
terms are generated.}
Typically this problem is overcome by adding in appropriate terms ``by hand,''
just as for DH theory \cite{fl,lf}.
In light of this fact, the conceptual advantage sometimes claimed for the
standard MSA in comparison to DH theory (see, e.g. [8(b)]), namely, that the
former treats the hard cores in better fashion, seems strictly inconsequential.
Note also that the density-density correlation functions, $G_{\rho \rho}({\bf
r})$, and also charge-charge correlation functions, $G_{q q}({\bf r})$, that
satisfy the Stillinger-Lovett second-moment-condition follow from DH theory
(again contrary to [8(b)]) when properly generalized \cite{benf}.

The pressure route to MSA thermodynamics (which we will denote MSpA) generates
a different approximation for the electrostatic excess free energy, along with
the Percus-Yevick-pressure-equation hard-core free energy.
It is interesting that, like the ordinary energy-route MSA thermodynamics, the
MSpA yields both a critical point and the exact DH limiting laws;
early on, however, Waisman and Lebowitz [11(c)] dismissed it as inferior.
By contrast, the compressibility route yields {\em no} electrostatic
contribution, but generates {\em only} the
Percus-Yevick-compressibility-equation free energy for uncharged hard spheres!
Finally, note that the thermodynamics of the generalized MSA or GMSA (which is
designed so that all three routes to the thermodynamics agree)
\cite{evans,gmsa} is identical to the ordinary, energy-route MSA combined with
the Carnahan-Starling (CS) approximation for the pure hard-core free energy
\cite{cs}.

\subsubsection{Hard Cores}
Since the RPM consists of hard spheres, it is certainly desirable to include an
account of the excluded volume effects in any approximate theory.
As we have seen, the two principal approximations, DH and MSA, require the
insertion of hard cores terms ``by hand,'' and two other theories, MSpA and
GMSA, entail two different hard-core approximations.
For the sake of convenience and uniformity, then, we will employ the CS hard
core approximation \cite{cs} in the calculations reported here for {\em all}
theories that recognize excluded volume effects.
The corresponding theories will be denoted DHCS, MSACS, and MSpACS, while the
notation DH, MSA, and MSpA will be reserved for the ``pure'' (electrostatics
only) theories.
We have, however, checked that other approximations for the pure hard-core
contributions yield qualitatively similar results.

It is worth mentioning that although hard-core terms do not contribute directly
to the internal energy (since their contribution to the energy of allowed
configurations vanishes --- as correctly reflected by the CS approximation),
they {\em do} influence the overall internal energy picture.
Specifically, for the basic theories, as we shall see, they affect internal
energy isotherms by altering the coexistence curve;
for the augmented, pairing theories, they enter by changing the degree of
pairing.

\subsection{Assessment of Basic Theories}
\subsubsection{DH Configurational Energy}
For pure DH theory (with neither pairing nor hard-core effects) the
configurational energy assumes a particularly simple form, namely,
\begin{equation}
\label{equdh}
u^{DH}(\rs,\ts) = -x/2(1+x).
\end{equation}
Evidently the energy of DH theory {\em violates none of the bounds} for any
values of $\rho$ and $T$: see (\ref{eqkapx}), (\ref{eqonsag}), and
(\ref{eqtot}).
Furthermore, $u^{DH}$ remains above the crystal values (\ref{eqxtal}) as is
apparent in Fig. 1.
The contrary statements by Blum and coworkers\cite{blum,bernard,blumnew} that
$u^{DH}$ violates Onsager's bound perhaps mistake the Debye-H\"{u}ckel limiting
law (DHLL) --- i.e., truncation of DH theory to lowest order in $x$, which no
one should take seriously for
$x \gtrsim 0.3$: see Fig. 1
--- for the full DH theory propounded in \cite{debye}.

Strictly, the dependence of $u^{DH}$ on the single parameter $x$ given in
(\ref{equdh}) can be correct only in single-phase regions of the ($\rho,T$)
plane.
Below the critical temperature (as defined by the theory at hand) the energy in
the coexistence region is always a weighted sum of the values in the two
phases, say $\al$ and $\be$.
In fact, if the energies per particle are $u_{\al}$ and $u_{\be}$ and the
densities $\rho_{\al} = \rho_{\al}(T)$ and $\rho_{\be} = \rho_{\be}(T)$,
one finds
\begin{equation}
\label{equcoex}
u(\rs,\ts)  = \frac{  \rho_{\al} (\rho_{\be} - \rho) u_{\al} + \rho_{\be} (
\rho - \rho_{\al}) u_{\be}  }{  \rho ( \rho_{\be} - \rho_{\al} )  } ,
\end{equation}
so that $u$ varies linearly with $1/\rho$.
Thus the main DH plot in Fig. 1 is restricted to $T \geq T_c^{DH}$, and
similarly for the other theories.
However, including phase coexistence according to (\ref{equcoex}) cannot induce
bound violation, since a weighted sum of two acceptable values also satisfies
the bound:
see the inset in Fig. 1 where the solid curves depict DH isotherms for $T \leq
T_c^{DH}$.

\begin{figure}
\narrowtext
\epsfxsize=\hsize
\epsfbox{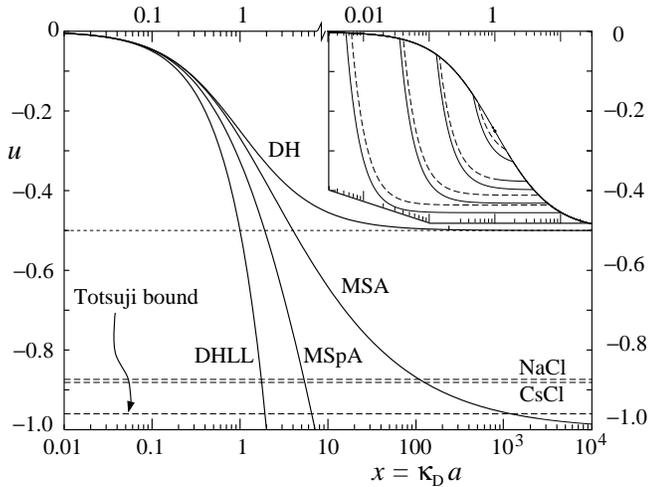}
\caption{The configurational energy per particle for the
Debye-H\"{u}ckel (DH), mean spherical approximation (MSA) and related theories
above criticality for comparison, with lower bounds.
For a description of the bounds and the theories, see the text.
The inset shows isotherms for $T \leq T_c$ for the DH and DHCS theories as
solid and dashed curves, respectively.
(Here and below, CS denotes use of the Carnahan-Starling approximation for the
excluded-volume effects.)}
\medskip
\end{figure}

Regarding the effects of hard cores, one finds that the only changes in DHCS
theory occur in the two-phase regions below $T_c^{DHCS}$: the energy isotherms
are shifted from those of pure DH theory since the coexistence curve differs.
The dashed curves in the inset to Fig. 1 show the rather small effects:
the shifts mainly reflect the expected lowered densities on the liquid branch
of the coexistence curve.
Naturally, these changes cannot induce any violation of Totsuji's bound or of
the crystal limits.

\subsubsection{MSA Configurational Energy}
Now Blum and Bernard \cite{blum,bernard} have claimed the energy of the (pure)
MSA, is ``asymptotically correct.''
However, as can be seen in Fig. 1, the MSA reduced excess energy, namely
\cite{waiseq},
\begin{equation}
\label{equmsa}
u^{MSA}(\rs,\ts)= - \left [ 1 + x - (1 + 2x)^{1/2} \right ]/x,
\end{equation}
asymptotically approaches the Onsager bound of $-1$ but {\em violates the
Totsuji bound} for $x \geq \xtee \simeq 1200$ (as Totsuji noted originally
\cite{tot}).
Furthermore, $u^{MSA}$ lies below the crystal values for $x \geq \xeks \simeq
125$.

In fact, even in the absence of Totsuji's result, it is hard to make sense of
the claim \cite{blum,bernard} that the MSA energy is asymptotically correct for
the RPM in the limit of large $x$ by virtue of its approach to Onsager's bound.
Agreement with a bound is hardly proof of correctness \cite{reblum}!
Furthermore, the limit $x \app \infty$ at fixed density implies $\ts \sim T/q^2
\app 0$;
but at low temperatures, one expects crystalline phases to appear for $\rs
\lesssim \rs_{max} = \sqrt{2}$ (for fcc sphere packing) \cite{fisher} and these
are not described by any of theories under consideration.

It is worthwhile to interpret more explicitly the values $\xtee$ and $\xeks$,
where violation by the pure MSA (no hard cores) occurs.
On the liquid side of the coexistence curve, $\xtee$ corresponds to violation
when $\ts \leq 0.012 \simeq (0.14) T^{*MSA}_c$ and $\xeks$ corresponds to $\ts
\leq 0.035 \simeq (0.41) T^{*MSA}_c$.
(The first violation temperature here is estimated with the aid of a
low-temperature asymptotic analysis of the pure MSA coexistence curve
\cite{msalowt} while the second follows directly from a numerical evaluation.)
The solid curves in Fig. 2 demonstrate the effects.

\begin{figure}
\narrowtext
\epsfxsize=\hsize
\epsfbox{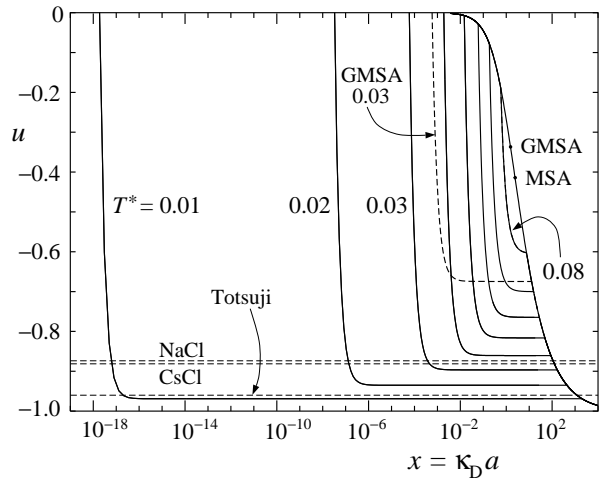}
\caption{Comparison of the MSA energy with bounds
for $T \leq T_c$ at multiples of $\ts = 0.01$ up to $\ts_c \simeq 0.0858$
(solid curves).
The dashed curve shows the $\ts = 0.03$ isotherm for the GMSA for which,
presumably, violations occur only at much lower temperatures.}
\end{figure}

The inclusion of hard-core terms (``by hand'') in the pure MSA changes the
liquid-side coexistence curves more strongly than in DH theory.
Thus for the MSA with CS terms or, equivalently, for the GMSA, the violations
shift to much lower ratios of $T/T_c^{GMSA}$: this is clearly evidenced by the
dashed coexistence isotherm shown in Fig. 2 for $\ts = 0.030 \simeq (0.38)
T_c^{*GMSA}$ (with $T_c^{*GMSA} \simeq 0.0786$ \cite{evans,tdgevans}).

\subsubsection{MSpA Configurational Energy}
The energy according to the MSpA is \cite{wais}
\begin{eqnarray}
u^{MSpA} = - \hbox{$\frac{1}{3}$} &[& 1 - ( 1 - \sqrt{1 + 2 x} )/x
                                \\
                               &+& 2 \ln{\left( 1 + x + \sqrt{1 + 2 x} \right)}
                               - 2 - \ln{4}] ,\nonumber
\end{eqnarray}
which, in the single-phase region, also depends only on the parameter $x$.
As evident from Fig. 1, however, this violates the Totsuji and Onsager bounds
at $\xtee \simeq 6.5$ and $\xoh \simeq 7.1$, respectively.
These results provide ample justification for a disparaging evaluation of the
pressure-route thermodynamics for the MSA.
For the remainder of this paper, we thus omit the MSpA.

\subsection{DH and MSA Free Energies}
\label{secsimgil}
In the pure theories (in which Bjerrum ion pairing is not explicitly included)
we find that {\em both} DH theory and the MSA {\em violate} Gillan's free
energy upper bound.
The entire vapor branches of both coexistence curves, as well as both sides of
the DH critical region, are in violation.
As shown in Fig. 3, the violations remain when hard-core excluded volume
corrections are included.
The DHCS and GMSA treatments exhibit very similar features, for the low
densities of interest.
Note that in Fig. 3 we follow the coexistence prescription for the free energy
corresponding to (\ref{equcoex}).
Note also that non-violation on one branch of the  coexistence curve (as on the
GMSA liquid side) is at best a qualified virtue since the construction of the
coexistence curve depends on the free energies on {\em both} sides.
In light of these results it is clearly imperative to examine theories which
allow for ion pairing.

\section{ASSESSMENT OF ION-PAIRING THEORIES}
\label{secaug}
\subsection{Bjerrum and Beyond}
To compensate for the effects of the DH linearization of the electrostatic
Boltzmann factor, Bjerrum \cite{bj} postulated association of ``free'' ions of
(residual) density $\rho_1$ into ``bound'' neutral dipolar pairs of density
$\rho_2$ so that the overall density is
\begin{equation}
\label{eqconserve}
\rho = \rho_1 + 2 \rho_2 .
\end{equation}
In terms of the ideal-gas free energy density
$\fbar^{Id}_j(\rho_j,T) = \rho_j [1 - \ln(\Lambda_j^{3j} \rho_j/\zeta_j)]$
with mean thermal de Broglie wavelengths $\Lambda_j(T)$ and internal partition
functions $\zeta_j(T)$ \cite{lf}, we may then write the total free energy
density as \cite{fl,lf}
\begin{equation}
\label{eqfbasic}
\fbar = 2 \fbar^{Id}_1(\hbox{$\frac{1}{2}$} \rho_1)
	+ \fbar^{Id}_2(\rho_2) + \fbar^{Ex}(\rho_1, \rho_2) ,
\end{equation}
with the excess free energy density
\begin{equation}
\label{eqfex}
\fbar^{Ex}(\rho_1, \rho_2) = \fbar^{HC}(\rho_1, \rho_2)
	+ \fbar^{Ion}(\rho_1) + \fbar^{DI}(\rho_1, \rho_2),
\end{equation}
where (i) $\fbar^{HC}$ denotes the pure hard-core/excluded-volume terms, (ii)
$\fbar^{Ion}$ represents the electrostatic contribution of the free ions, while
(iii) $\fbar^{DI}$ denotes the dipole-ion interaction/solvation terms
\cite{fl,lf}.
As mentioned, we take here $\fbar^{HC}$ to be of Carnahan-Starling form
\cite{cs} with the dipoles treated as effective spheres of diameter $\sigma_2 =
2^{1/3} a$ \cite{benf}.

\begin{figure}
\narrowtext
\epsfxsize=\hsize
\epsfbox{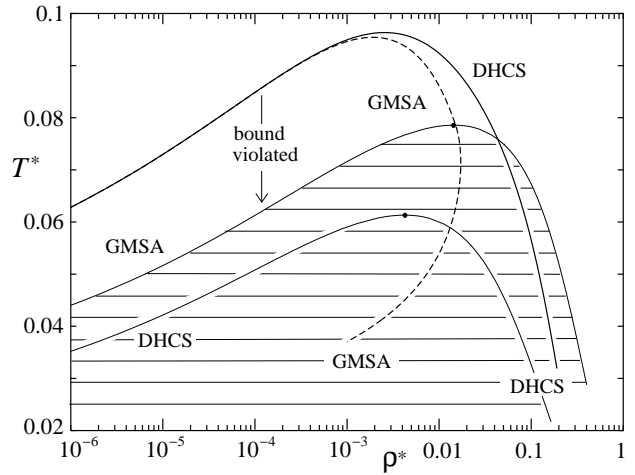}
\caption{Comparison of
the free energies predicted by the DHCS and GMSA theories
in the density-temperature plane with
Gillan's upper bound.
The bound is violated below the solid and dashed curves, respectively.
For comparison, the associated coexistence curves with tie-lines and critical
points are also plotted.}
\end{figure}

Chemical equilibrium among the $+$ and $-$ free ions and dipolar pairs is
imposed via the equality $\mu_2 = 2 \mu_1$ of the chemical potentials.
If the association constant is defined by
$K(T) = \Lambda_+^3 \Lambda_-^3 \zeta_2 / \zeta_+ \zeta_-\Lambda_2^6 = \zeta_2$
(see \cite{lf}) and the reduced excess chemical potentials are
\begin{equation}
\label{eqchempot}
\overline{\mu}^{Ex}_j \equiv \mu^{Ex}_j/k_B T
= \ln{\gamma_j}
= - (\dee \fbar^{Ex}/ \dee \rho_j),
\end{equation}
with $\rho_+ = \rho_- = \frac{1}{2} \rho_1$ and  $\gamma_+ = \gamma_- =
\gamma_1$, then the mass action law states
\begin{equation}
\label{eqmassaction}
\frac{\rho_2}{\rho_+ \rho_-} = \tilde{K}(T; \rho_1, \rho_2)
\equiv K(T) \frac{\gamma_+ \gamma_-}{\gamma_2}.
\end{equation}
The optimal expression for $K(T)$ is a matter for debate \cite{fl,lf} --- and
will be discussed further below.
For reference purposes we adopt Ebeling's form \cite{lf,ebgr,ebeling} which
guarantees an exact representation of the RPM's electrostatic second virial
coefficient when one uses DH theory or the MSA (but not the MSpA) for
$\fbar^{Ion}(\rho_1)$.
Note that for $\ts \leq 0.05 \simeq \ts_c$ the difference between $K^{Eb}$ and
Bjerrum's original proposal, $K^{Bj}$, is less than 0.01\%; it rises to 3.0\%
at $\ts = T_c^{*MSA} = 0.0858$, in accord with the Introduction.

Bjerrum's original theory \cite{bj} amounts to the approximation
\begin{equation}
\mbox{DHBj:} \;\;\;
\fbar^{Ex} \simeq \fbar^{Ion} \simeq \fbar^{DH}(x_1) \;\;\;\; \mbox{with}
\;\;\;\; x_1 = \kappa_1 a ,
\end{equation}
where $\kappa_1^2 = 4 \pi q^2 \rho_1/D k_B T$ represents the inverse squared
Debye length for the {\em free ions alone}, while as usual \cite{debye},
\begin{equation}
\label{eqfdh}
\fbar^{DH}(x) =
\left [ \ln{(1 + x)} - x + \hbox{ $\frac{1}{2}$ } x^2  \right ]/ 4 \pi a^3.
\end{equation}
Friedman and Larsen \cite{friedlar} later found that the predicted coexistence
curve was unphysical.
More recently, Fisher and Levin \cite{fisher,fl,lf} elucidated the peculiar
``banana'' shape of the DHBj coexistence curve (see Fig. 4 below) and showed it
became worse when excluded-volume terms were added as, e.g., in DHBjCS theory.
However, they also estimated the dipole-ion solvation term as \cite{lf}
\begin{equation}
\label{eqfdi}
\fbar^{DI} = \rho_2 (a a_1^2/a_2^3 \ts) \tilde{\omega}_2(x_2), \;\;\;\;\;
	x_2 = \kappa_1 a_2,
\end{equation}
\begin{equation}
\label{eqdiomega}
\tilde{\omega}_2(x) = 3 \left [ \ln{( 1 + x + \hbox{$\frac{1}{3}$} x^2 )}
	- x + \hbox{$\frac{1}{6}$} x^2 \right ]/x^2 \approx x^2/12,
\end{equation}
where $a_1 = (1.0 \mbox{-} 1.3)a$ is the mean dipolar size, or $+/-$ ion
separation, while $a_2 \simeq 1.1619_8 a$ represents the effective
electrostatic exclusion radius \cite{lf}.
(Note that all the results given here use $a_1 = a$ and $a_2 = 1.16198 a$.)
The resulting DHBjDI theories lead to sensible coexistence curves (see Fig. 5
below) that agree fairly well with current simulations [5,2(b)].

\begin{figure}
\narrowtext
\epsfxsize=\hsize
\epsfbox{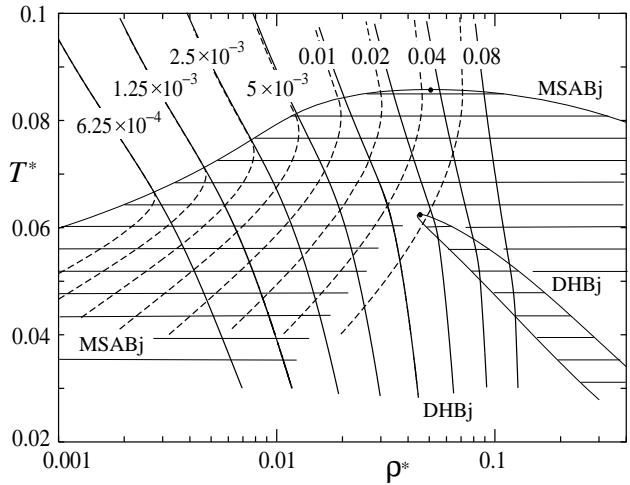}
\caption{Pure 
Bjerrum pairing theories tested against Gillan's free-energy bound.
The solid and dashed ``excess contours'' are labeled by the magnitudes by which
the DHBj and MSABj reduced free energies, respectively, fall below the upper
bound (see text).
Note the associated coexistence curves and the unrealistic ``banana'' shape of
the DHBj prediction \protect\cite{fisher,fl,friedlar}.}
\end{figure}

At an earlier stage, Ebeling and Grigo \cite{ebgr} combined Bjerrum pairing
with the MSA by replacing $\fbar^{DH}$ by \cite{evans,wais}
\begin{equation}
\label{eqfmsa}
\fbar^{MSA}(x) = \left[ 2 + 6x + 3x^2 - 2 (1 + 2x)^{3/2} \right ]/12 \pi a^3,
\end{equation}
with $x \Rightarrow x_1$ again evaluated at $\rho_1$.
They also added excluded-volume terms.
The resulting MSABj and MSABjCS $\equiv$ EGA [8(b),14] theories yield fully
acceptable coexistence curves \cite{lf} but, as mentioned, the predicted
critical temperatures are significantly too high [2(b),5].

\begin{figure}
\narrowtext
\epsfxsize=\hsize
\epsfbox{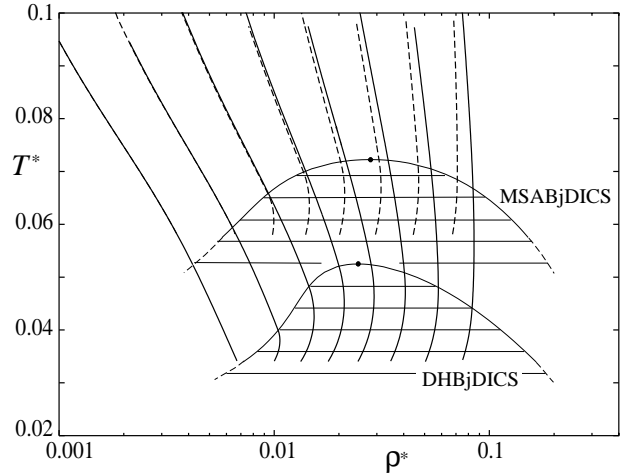}
\caption{Comparison 
of BjDICS free energies, which incorporate dipole-ion solvation and
Carnahan-Starling excluded-volume terms, with the Gillan bound, as in Fig. 4.}
\end{figure}

Recently, Zhou, Yeh, and Stell (ZYS) \cite{stellassoc} have extended Ebeling's
approach by using the MSA in conjunction with a ``reference cavity theory of
association'' \cite{zhoustell}.
Their {\em pairing mean-spherical approximations} or PMSA theories may be
described by
\begin{eqnarray}
\mbox{PMSA:}&& \;\;\; \nonumber\\
\fbar^{Ex} &=& \fbar^{MSA}(x) + \fbar^{CS}(\rho)
	+ \rho_2(T,\rho) \ln (\gamma_+ \gamma_- / \gamma_2),\nonumber\\
&&\label{eqpmsagen}
\end{eqnarray}
where $x = \kappa_D a$ is now evaluated with the {\em total density}, $\rho$,
and $\fbar^{CS}$ represents the single-component Carnahan-Starling form,
evaluated at $\rho = \rho_1 + 2 \rho_2$ (i.e., bound pairs are not treated as
geometrically distinct objects).
Note that $\rho_2$ is here to be determined from (\ref{eqmassaction}) once $K$,
$\gamma_1$, and $\gamma_2$ are specified (see below);
hence $\rho_2$ is an explicit algebraic function of the arguments stated in
(\ref{eqpmsagen}).
The use of only the total density (in place of the free ion density $\rho_1$)
results in an analytically simpler, more explicit formulation; but, in the
light of the original DH and Bjerrum arguments, it seems rather unphysical
since neutral bound pairs cannot contribute to screening in a direct way.
Furthermore, as we will see, this approach entails a significant cost in
accuracy.

The specification of the PMSA may be completed by first noting that ZYS also
adopt Ebeling's association constant, $K^{Eb}(T)$ \cite{lf,ebgr,ebeling}.
Then, for the activity coefficients, $\gamma_+ \equiv \gamma_-$ and $\gamma_2$,
ZYS propose three levels of approximation, first:
\begin{eqnarray}
\mbox{PMSA1:} &&\nonumber\\
\ln{\gamma_1} &=& -(\partial \fbar^{MSA}/ \partial \rho )_T
\equiv \overline{\mu}^{MSA}(T,\rho),  \;\;\; \gamma_2 = 1,\nonumber\\
&&
\end{eqnarray}
which neglects dipole-ion contributions [cf. (\ref{eqpmsagen})].
Second, dipole-ion interactions are introduced by replacing the approximation
$\gamma_2 = 1$ by
\begin{eqnarray}
\mbox{PMSA2:}&&\nonumber\\
\ln{\gamma_2} & = & \left [ 2(1+x)\sqrt{1+2x} - 2 - 4x - x^2 \right ]/ \ts x^2,
\nonumber \;\;\;\;\;\;\;\;\;\; \\
	& \approx & -x^2/4 \ts [1 + O(x)];
\end{eqnarray}
see [8(b)], Eq.~(4.11).
Finally, the dumbbell-shaped hard cores of a dipolar ion pair are incorporated
[8(a)] by using the CS cavity-value contact function and incrementing
$\ln{\gamma_2}$ by
\begin{equation}
\label{eqpmsa3}
\mbox{PMSA3:} \;\;\;\;\;
\Delta \ln{\gamma_2} = \ln{[ 2(1-\eta)^3/(2-\eta) ]},
\end{equation}
where $\eta = \pi \rs / 6$.

PMSA3 is the preferred theory of ZYS and yields $(\ts_c,\rs_c) \simeq
(0.0745,0.0245)$.
PMSA1 and PMSA2 give $(0.0748,0.0250)$ and $(0.0733,0.0229)$, respectively.
The $T_c$ values are still significantly higher [8(b)] than the DH-based
estimates, namely, $\ts_c \simeq 0.052 \mbox{-} 0.057$
\cite{fisher,lf,fishlee}, while the simulations suggest $\ts_c \simeq 0.048
\mbox{-} 0.055$ [2(b),15].

\subsection{Pairing Theories vs. Gillan's Bound}
Comparison of the pairing theories with Gillan's free energy bound is mainly
encouraging.
We find that theories that incorporate association in the Bjerrum chemical
picture, in which the free ion density is {\em depleted} by pairing (i.e.,
$\rho_1 = \rho - 2 \rho_2$), never violate the bound.
Indeed, even the most primitive Bjerrum theories, DHBj and MSABj --- which
include neither hard-core nor dipole-ion interactions --- satisfy Gillan's
bound for all $(\rs,\ts)$ values tested: see Fig. 4.
On the other hand, all three PMSA theories turn out to violate Gillan's bound
in significant regions of the $(\rs,\ts)$ plane, including nearly the entire
vapor branches of the coexistence curves.

As regards the MSABj and DHBj theories, the more-or-less vertical ``excess
contour lines'' in Fig. 4 reveal the magnitude of non-violation in the
density-temperature plane:
they are loci on which Gillan's upper bound exceeds the corresponding
approximate reduced free energy density, $-\fbar a^3$, by the indicated
amounts, ranging from $6 \times 10^{-4}$ up to $0.1$.
The associated coexistence curves are also shown and one may notice that the
excess contours undergo a jump in curvature on entering the corresponding
two-phase region: this results from the coexistence prescription analogous to
(\ref{equcoex}).

Fig. 5 shows the effects of incorporating dipole-ion solvation (DI) and
excluded-volume (CS) terms.
Note that removing the excluded-volume terms from these BjDICS theories
produces only slight shifts in the excess contours at high densities and low
temperatures.

By contrast, the solid curve in Fig. 6 marks the boundary of the region inside
which the PMSA3 free energy violates Gillan's bound.
The coexistence curve is also shown.
(Note, however, that the coexistence prescription was not used here to compute
the violation boundary within the two-phase domain.)
The region of violation found for PMSA2 is nearly identical, while that for the
PMSA1 theory is slightly larger, extending {\em above} the corresponding
critical point, $T_c^{PMSA1}$: see the dashed curve in Fig. 6.

\begin{figure}
\narrowtext
\epsfxsize=\hsize
\epsfbox{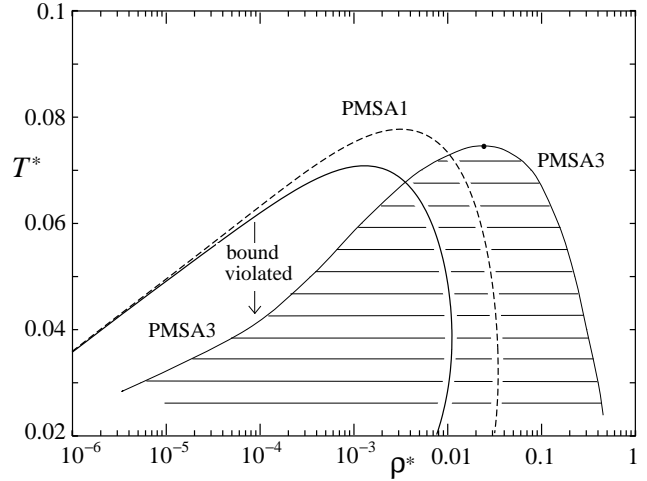}
\caption{Test of the PMSA theories against Gillan's free energy bound.
All theories fail at low temperatures and densities:
see the violation boundaries, solid for PMSA3 (the preferred theory) and dashed
for PMSA1.
The coexistence curve and critical point are those predicted by the PMSA3.}
\end{figure}

In conclusion, the violations of Gillan's bound found previously and seen here
for the PMSA theories demonstrate convincingly that association of oppositely
charged ions into dipoles {\em along with} a concomitant depletion of free ions
and their screening effects is a crucial element in the critical-region
behavior of the RPM.
Gillan's bound also serves to highlight interesting contrasts between DH- and
MSA-based theories:
the MSA coexistence curve shifts only slightly when pairing is added (MSABj)
yet, surprisingly, violation of Gillan's bound is still completely avoided;
the unphysical DHBj ``banana'' coexistence curve (in Fig. 4), on the other
hand, immediately points to the significance of pairing, while satisfaction of
Gillan's bound is surprising here because the coexistence curve is so
unconvincing.

\subsection{Pairing Theories vs. Energy Bounds}
Testing the pairing theories against the bounds of Totsuji and Onsager yields
mixed results.
For a window of temperatures that includes the critical region, namely, $0.015
\lesssim \ts \lesssim 0.5$, all the theories embodying ion association satisfy
the energy bounds.
We also find a surprising level of agreement among the various theories as to
the value of the critical energy per particle: see Table I.
At low temperatures, however, some of the MSA-based theories violate Totsuji's
bound.
Moreover, at moderate temperatures ($\ts \gtrsim 0.5$) {\em all} of the pairing
theories violate fundamental thermal stability requirements (as discussed in
the next section); for some of the approximations, this is also accompanied by
violation of the Totsuji and Onsager bounds, as explained below.

Now the energy for a general pairing theory follows from (\ref{eqfbasic}) via
the thermodynamic relation (\ref{equdef}) and the mass action law
(\ref{eqmassaction}), etc., which leads to
\begin{equation}
\label{equwithpair}
u(\rho,T) =
	\frac{a^3 T^{\ast 2}}{\rs} \frac{\dee}{\dee \ts} \fbar^{Ex}(\rho,T)
	+ \frac{\rho_2}{\rho} u_2(T),
\end{equation}
where $u_2(T)$ is given by
\begin{equation}
\label{equ2}
u_{2}(T) = T^{\ast2} (d \ln{K(T)}/d \ts )
\end{equation}
But this can be recognized simply as the mean energy of a {\em single} $(+,-)$
bound pair since the corresponding internal configurational partition function
for a pair is embodied in the association constant, $K(T)$ --- see the text
above (\ref{eqchempot}) and Levin and Fisher \cite{lf}.
Of course, the factor $\rho_2(\rho,T)$ in (\ref{equwithpair}) is also to be
determined via the law of mass action (\ref{eqmassaction}).
For theories of the form (\ref{eqfex}), one can further write
\begin{equation}
\label{equex}
u^{Ex} = (a^3 T^{\ast 2}/\rs) (\dee \fbar^{Ex}/ \dee \ts)
	= u^{Ion} + u^{DI},
\end{equation}
where the ``basic'' expressions for the electrostatic contribution, $u^{Ion}$,
are now given by the natural generalizations of (\ref{equdh}) and
(\ref{equmsa}), namely,
\begin{equation}
\label{equdhpair}
u^{DH}(\rho,T) = -(\rho_1/\rho)x_1/2(1+x_1),
\end{equation}
\begin{equation}
\label{equmsapair}
u^{MSA}(\rho,T)= - (\rho_1/\rho)
	\left [ 1 + x_1 - (1 + 2x_1)^{1/2} \right ]/x_1 .
\end{equation}
For reference, we also quote the explicit result for $u^{DI}$ following from
the treatment of Fisher and Levin in leading order \cite{DInote}.
Defining $a_1$ and $a_2$ as in (\ref{eqfdi}) and (\ref{eqdiomega}) \cite{lf},
one finds
\begin{equation}
u^{DI} =
	- \frac{aa_1^2}{2 a_2^3}\frac{\rho_2}{\rho}
               \frac{(\xt)^2}{ [3 + 3 \xt + (\xt)^2] } .
\end{equation}
The corresponding expressions for the PMSA theories are omitted for the sake of
brevity.

\end{multicols}
\begin{table}
\widetext
\caption{Some critical-point parameters for various theories:
$\ts_c$; $u_c$, the reduced energy per particle;
$x_c = (4 \pi \rs_c / \ts_c)^{1/2}$, the (overall) Debye parameter; and,
$x_{1c} = (4 \pi \rho^*_{1c} / \ts_c)^{1/2}$, the screening parameter.
(Note that the values quoted for $x_c$ in \protect\cite{lf} correspond 
here to $x_{1c}$ and that the Ebeling association constant 
\protect\cite{ebgr} was used throughout.)}
\medskip
\begin{tabular}{c|cccccc}
$ \;\;\;\;\;\;\;\;\;\; $ &DH      &$+$CS      &$+$Bj      &$+$BjCS    &$+$BjDI
  &$+$BjDICS \\[0.1cm] \hline 
$\ts_c$   &$0.0625$   &$0.061_3$  &$0.0625$   &$0.061_5$  &$0.057_4$
&$0.052_5$ \\
$u_c$	  &$-0.25$    &$-0.241_1$ &$-0.431_5$ &$-0.437_8$ &$-0.444_3$ &$-0.453_3$
\\
$x_c$     &$1$        &$0.931_5$  &$3.013_5$  &$3.281_1$  &$2.466_1$
&$2.424_0$ \\
$x_{1c}$  &$1$        &$0.931_5$  &$1$        &$0.938_6$  &$1.122_9$
&$0.931_5$ \\[0.1cm]
\hline \hline
	  &MSA	      &$+$CS      &$+$Bj      &$+$BjCS    &$+$BjDI    &$+$BjDICS \\
\hline
$\ts_c$   &$0.085_8$  &$0.078_6$  &$0.085_8$  &$0.078_7$  &$0.082_1$
&$0.072_3$ \\
$u_c$     &$\;\;\;-0.414_2\;\;\;$ &$\;\;\;-0.335_8\;\;\;$
&$\;\;\;-0.415_7\;\;\;$ &$\;\;\;-0.378_1\;\;\;$ &$\;\;\;-0.444_2\;\;\;$
&$\;\;\;-0.414_8\;\;\;$ \\
$x_c$     &$2.414_2$  &$1.522_1$  &$2.721_3$  &$2.040_8$  &$3.072_9$
&$2.208_3$ \\
$x_{1c}$  &$2.414_2$  &$1.522_1$  &$2.414_2$  &$1.531_9$  &$2.450_9$
&$1.485_0$ \\[0.1cm]
\hline \hline
	  & PMSA1     & PMSA2     & PMSA3     &           &           &          \\
\hline
$\ts_c$   &$0.073_3$  &$0.074_8$  &$0.074_5$  &           &           & \\
$u_c$     &$-0.374_0$ &$-0.426_6$ &$-0.426_5$ &           &           & \\
$x_c$     &$1.981_4$  &$2.049_4$  &$2.032_9$  &           &           & \\
\end{tabular}
\end{table}
\begin{multicols}{2}

\subsubsection{Low Temperatures: Violation in MSA Pairing Theories}
For $\ts \lesssim 0.015$, evaluation of $u(\rho,T)$ reveals violations of the
Totsuji bound for most of the MSA theories.
The reason turns out to be literally the same as for the pure MSA: in the
corresponding Bj, BjCS, BjDI, and BjDICS theories, as well as in the PMSA1
(although {\em not} PMSA2 and 3) theory, the mass-action pairing predicted by
(\ref{eqmassaction}) becomes exponentially small as $\ts \app 0$
\cite{lfmsapair}.
As a result, all these theories revert to their ion-only form (i.e., MSA or
MSACS) and violations occur: see Fig. 1.
A similar depletion of pairs occurs when $\ts \app 0$ in the DHBjDI/CS (but
{\em not} DHBj/CS) theories, and so these theories revert to the corresponding
{\em non-violating} DH/CS theories.
These results are independent of whether one uses the Ebeling or Bjerrum
association constant, or any other reasonable partition-function-like form, as
discussed below.

\subsubsection{Moderately Low Temperatures}
In the temperature range $0.015 \lesssim \ts \lesssim 0.5$, which includes
$\ts_c$, all the pairing theories described in the present study satisfy the
Totsuji bound, and hence Onsager's as well.
Fig. 7 depicts energy isotherms for the pairing theories at $\ts = 0.07$ .
The plotted isotherms have been cut off for large $x = \kappa_D a$ at the
hard-core packing limit, $\rho^{\ast}_{max} = \sqrt{2}$.
Fig. 7 also shows the location of the critical point of the DHBjDICS theory,
which may be regarded as a reference point in reading Table I.
The table lists the various critical energies and Debye parameters.
As mentioned, there is a fair measure of agreement among the different pairing
theories regarding the energy at criticality even though other parameters vary
quite strongly.

\begin{figure}
\narrowtext
\epsfxsize=\hsize
\epsfbox{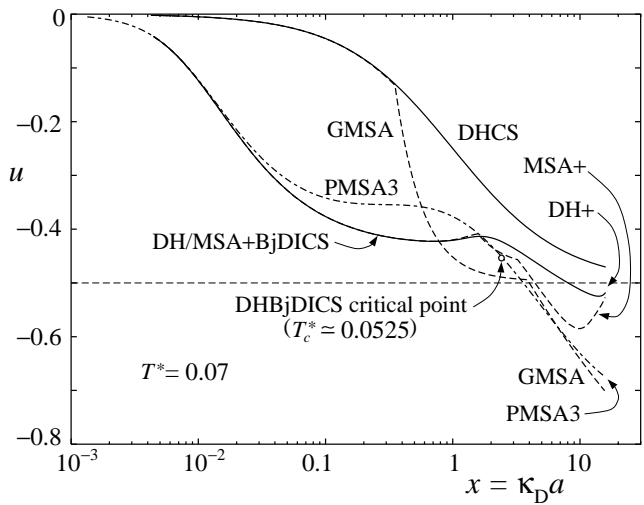}
\caption{Plots 
of configurational energy isotherms for various theories at $\ts = 0.07$,
which temperature lies {\em above} all DH-based estimates of $\ts_c$ but {\em
below} all MSA-based values.
The scalloped sections of the latter isotherms thus represent the two-phase
regions.
As regards the theories, recall that GMSA is equivalent to MSACS and note that
the ``$+$'' implies the BjDICS extensions of the basic theories.
At large $x = \kappa_D a$ all isotherms have been cut off at the hard-sphere
close packing density.
For reference, the critical point of the DHBjDICS theory (where $\ts_c =
0.052_5$) has been marked.}
\end{figure}

\subsubsection{Violations at Moderate and High Temperatures}
Violation of the energy bounds are found again, as mentioned above, at {\em
higher} temperatures in the range $\ts \gtrsim 0.5$, some 6 to 10 times greater
than the estimates for $\ts_c$.
The reason for this surprising fact, however, is quite different from the cause
at low temperatures:
it transpires, indeed, that the form of the association constant is now
crucially important.

In fact, any theory with pairing governed by Bjerrum's association constant
violates both the Totsuji and Onsager bounds when $\ts \app \frac{1}{2}-$ and
$\rho$ is large enough!
Once noticed numerically, this behavior can be understood analytically by
evaluating the factor $u_2(T)$ in (\ref{equwithpair}) using (\ref{equ2}) with
$K = K^{Bj}(T)$.
To that end recall, first, the well known fact \cite{lf} that $\kbjt$ vanishes
linearly, say as $c_{Bj}(1 - 2 \ts)$, when $\ts \app \frac{1}{2}-$ (and remains
identically zero for $\ts > \frac{1}{2}$).
Consequently, $u_2(T)$ diverges to $-\infty$ like $-\frac{1}{2}/(1 - 2 \ts)$ in
this limit.
However, the factor $\rho_2(\rho,T)$ in (\ref{equwithpair}) must be evaluated
via the mass action law (\ref{eqmassaction}) and is proportional to $\kbjt$;
this gives
\begin{equation}
\label{equpairterm}
\frac{\rho_2}{\rho} u_2
= \frac{\rho_1^2 \gamma_1^2}{4 \rho \gamma_2} T^{*2} \frac{dK}{d\ts}
\approx -\frac{c_{Bj}}{8 a^3} \frac{\gamma_1^2}{\gamma_2} \rs < 0,
\end{equation}
as $\ts \app \frac{1}{2}-$, so that $\rho_2 \app 0$ and $\rho_1 \app \rho$.
Note that the $\gamma_i(\rho,T)$, defined via (\ref{eqchempot}), depend on the
theory under consideration.
One finds that $c_{Bj}/8 a^3 \simeq 11.6$: this is large enough so that the
pairing term (\ref{equpairterm}) by itself yields a violation of Onsager's
bound when (in DHBj theory) $\rs > \rho_{\mbox{\scriptsize
Ons}}^{*\mbox{\scriptsize DHBj}} \simeq 0.39$ or (for MSABj) $\rs >
\rho_{\mbox{\scriptsize Ons}}^{*\mbox{\scriptsize MSABj}} \simeq 0.64$.
However, as the other terms in (\ref{equwithpair}) are also negative,
violations must arise at even lower densities.
One finds numerically, in fact, that the violations occur at or below $\rs
\lesssim 0.3$ in {\em all} the theories with pairing governed by Bjerrum's
association constant.

One expects Ebeling's choice, $\kebt$, which provides a match to the exact RPM
second virial coefficient and never vanishes \cite{lf,ebgr,ebeling} --- in
contrast to the singular vanishing of $\kbjt$ at $\ts \geq \frac{1}{2}$ --- to
fare better.
Nevertheless, Ebeling's association constant leads to Onsager and Totsuji bound
violations in the region $\ts \simeq 0.7 \, \mbox{-} \, 1.0$ --- although only
in those theories which explicitly allow for the excluded volume effects.
The PMSA3 treatment, furthermore, falls into this same category of violation;
however, PMSA1 and 2 do {\em not} because the excluded-volume terms there do
not affect the degree of pairing.

All the violations just described turn out to be symptoms of a more serious
weakness of both the Bjerrum and Ebeling association constants, as we will now
demonstrate.

\subsection{Violations of Thermal Stability}
To pursue further the origins of the Totsuji and Onsager energy bound
violations at $\ts \gtrsim 0.5$, consider the energy isochores shown in Fig. 8.
The two densities $\rs = 0.03$ and $0.1$ have been chosen for display because
they bracket the critical density;
similar behavior is seen at higher and lower densities.
For the pure DH and MSA theories, included in Fig. 8 for reference purposes,
$u(\rho,T)$ rises monotonically with $T$:
this implies a positive constant-volume configurational specific heat,
$C_V^{\mbox{\scriptsize conf}}(\rho,T)$.
(Note that outside the two-phase region these two energy isochores are
identical to those for DHCS and GMSA, respectively.)

Now the positivity of the {\em total} constant-volume specific heat is a
thermodynamic necessity dictated by the Second Law \cite{pippard}.
For a classical particle system, however, the configurational contribution must
be separately nonnegative:
this follows {\em either}, thermodynamically, by regarding the kinetic and
configurational degrees of freedom as thermally distinct systems {\em or}, from
statistical mechanics, by expressing $C_V^{\mbox{ \scriptsize conf}}(\rho,T)$
as a mean-square energy fluctuation which is necessarily positive at finite
positive temperatures in any nontrivial system \cite{cvcon}.

\begin{figure}
\narrowtext
\epsfxsize=\hsize
\epsfbox{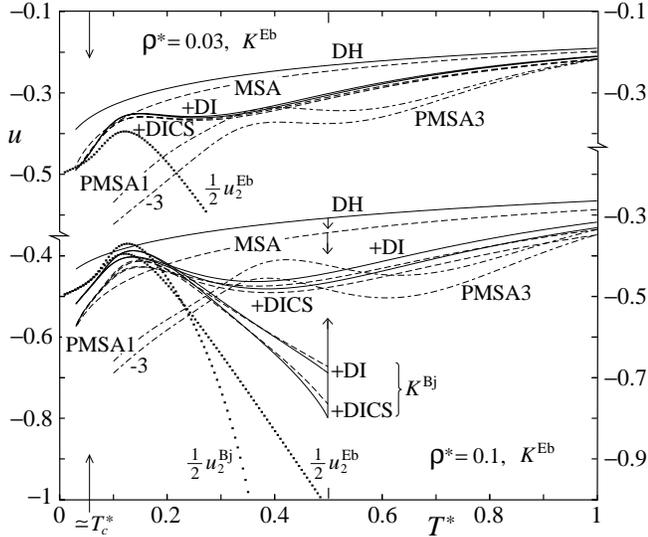}
\caption{Energy 
isochores at $\rs = 0.03$ and $0.1$ (note shifted vertical scales) for
the basic DH and MSA theories and for various pairing theories --- solid lines
for those based on DH, dashed lines for MSA-based.
Ebeling's association constant is employed for all plots excepting the four
bracketed isochores for $\rs = 0.1$ labeled $\kbj$, which use Bjerrum's
expression (which vanishes at $\ts = \half$).
The PMSA isochores are shown as dot-dash curves.
The plots labeled $\half u_2^{Bj}$ and  $\half u_2^{Eb}$ represent {\em
complete} ion pair association ($\rho_1 = 0, \; \rho_2 = \half \rho$), while
$u_2^{Bj}$ and  $u_2^{Eb}$ are corresponding single-pair energies implied by
the mass-action law.
Except for these plots, the isochores have been cut off below $\ts = 0.03$
because by then the extrapolation below $\ts_c$ into the two-phase regions
loses all significance.
[Note that, for the approximations considered here, $u^{DH}(\rho,T) =
u^{DHCS}(\rho,T)$ and $u^{MSA}(\rho,T) = u^{GMSA}(\rho,T)$ {\em outside} the
respective two-phase regions (see Sec. IIIA.3).]}
\end{figure}

However, a quick perusal of Fig. 8 shows that {\em all} the pairing theory
isochores --- the solid and dashed curves representing DH- and MSA-based
theories, respectively, and the dot-dash plots for PMSA1 and 3 --- display
regions where $u(\rho,T)$ {\em decreases} as $T$ increases.
In other words, all the paring theories predict negative constant-volume
specific heats and violate the Second Law!

The reason is not far to seek.
In the limit of complete pairing (i.e., $\rho_1 = 0, \; \rho_2 = \frac{1}{2}
\rho$) all the approximate theories under consideration predict, via
(\ref{equwithpair}), that the energy should be simply that of independent
dipolar pairs:
this corresponds to the plots labeled $\half u_2^{Bj}$ and $\half u_2^{Eb}$ in
Fig. 8 which derive from (\ref{equ2}) and the Bjerrum and Ebeling forms for
$K(T)$.
But, as evident from the figure, both $u_2^{Bj}(T)$ and $u_2^{Eb}(T)$ exhibit
pronounced maxima in the interval $\ts = 0.12 \, \mbox{-} \, 0.13$ and then
fall sharply as $T$ increases, dropping below $u_2^{Bj}(0) = u_2^{Eb}(0) = -1$
at $\ts = 0.22_2$ and $0.21_9$, respectively.
It is this behavior that leads to the decreasing regions in the overall excess
energy isochores with incomplete pairing.
But such a variation of $u_2(T)$ is physically nonsensical since, clearly, the
configurational energy $\etwo(r) = -q^2/Dr$ of a bound pair cannot fall below
the contact value $-q^2/Da$ (which, in turn, can be achieved in equilibrium
only at $T=0$).

The problem with $u_2(T)$ arises because the defining relation (\ref{equ2})
does not actually yield the physically anticipated thermodynamic mean value
\cite{halpern}, say $\langle \etwo(r) \rangle_K$, which in the Bjerrum picture
of association would be
\begin{equation}
\label{eqetwoav}
\langle \etwo(r) \rangle_K =
4 \pi \int_a^R \etwo(r)e^{-\beta \etwo(r)} r^2 dr / K(T),
\end{equation}
with association constant
\begin{equation}
\label{eqkoft}
K(T) = 4 \pi \int_a^R e^{-\beta \etwo(r)} r^2 dr .
\end{equation}
The reason for the failure is simple:
the Bjerrum cutoff $R$ is taken to be temperature dependent \cite{halpern},
explicitly, $R^{Bj}(T) = a/2 \ts$ for $\ts \leq \half$ \cite{lf,bj}.
In general, such temperature dependence leads to the difference
\begin{equation}
\label{equ2diff}
\frac{q^2}{Da} u_2(T) - \langle \etwo \rangle_K
= \frac{ 4 \pi R^2 e^{-a/\ts R} }{K(T)} k_B T^2 \frac{dR}{dT},
\end{equation}
which is negative whenever $R(T)$ decreases as $T$ rises and which diverges
when $K(T) \app 0$.
The Ebeling association constant can also be written in the form (\ref{eqkoft})
but with the large-$T$ asymptotic form $R^{Eb}(T) - a \approx a/12 T^{*4}$
\cite{lf}, which is quite accurate once $\ts \gtrsim 0.3$.
We must conclude that neither the Ebeling nor the Bjerrum association constants
can be regarded as representing even an ``effective'' partition function for an
isolated ion pair as is required by or implicitly assumed in the standard
theories of association \cite{lf,davidson}.

As suggested by Fig. 8, the unphysically large values of $u_2(T)$ lead to
negative specific heats over large regions of the $(\rho,T)$ plane when either
$\kebt$ or $\kbjt$ is employed.
Fortunately for our primary focus on the critical region, the violations of
thermal stability are confined in all cases to $\ts \geq 0.12 > 2 \ts_c$ (and
for the PMSA theories to $\ts \gtrsim 0.35$).
At densities below $\rs = 0.01 \, \mbox{-} \, 0.02 < 0.6 \rs_c$ the pairing is
sufficiently weak that the predicted $\cvcon(\rho,T)$ always remains positive
--- although it does display an unphysical oscillation.
Once violations arise at a given $T$, moreover, they persist to the highest
densities.

Of course, certain features are specific to the choice of association constant.
As remarked earlier, $\kbjt$ ``switches off'' abruptly at $\ts = \half$ where a
nonphysical latent heat is implied for all $\rho > 0$; above $\ts = \half$
pairing is lost and no violations remain.
When $\kebt$ is used in DH- and MSA-based theories with excluded-volume terms,
violations remain at the highest temperatures.

What might be a cure for these pathologies?
It is clear from (\ref{eqetwoav})-(\ref{equ2diff}) that the unphysical behavior
of $u_2(T)$ can be avoided if one fixes the cutoff in (\ref{eqkoft}) at, say $R
= \lambda a$, so defining $K^{\lambda}(T)$.
Furthermore, for any fixed $\lambda > 1$, the $\mbox{low-}T$ behavior of
$K^{\lambda}(T)$ still matches $\kebt$ to all orders in $\ts$ \cite{lf,fuoss}.
In addition, the choice of $\lambda$ may be optimized by requiring that the
deviation $|(\keb/K^\lambda) - 1| \equiv \delta$ remain less than a specified
level up to as high a temperature as possible.
Thus one finds that $\lambda \simeq 3.4$ provides 1\% precision ($\delta =
0.01$) up to $\ts \simeq 0.11$.

One can then check that {\em none} of the pairing theories employing
$K^\lambda(T)$ with $\lambda \simeq 3.4$ violates the energy bounds or thermal
stability for any realizable thermodynamic state, $(\rho,T)$.
In addition, the qualitative conclusions regarding the violation and
nonviolation of the Gillan free energy bound remain unchanged.
Indeed, using $K^{3.4}(T)$ causes only insignificant shifts of the free energy
excess contours from those displayed in Figs. 4 and 5 when $\ts \lesssim 0.1$.

Nevertheless, merely replacing $\kebt$ by $K^\lambda(T)$ leads to significant
inaccuracies in the thermodynamics at {\em higher} temperatures.
Thus a more thoughtful approach is essential to providing a reasonable
approximate theory of the RPM that is valid over the full range of temperatures
[and up to moderate densities excluding, of course, the solid phase(s)].
Such a treatment will be presented elsewhere \cite{assocpaper}.

\acknowledgments

The second author is grateful for the stimulus provided for this work by
Professor Joel L. Lebowitz and by his attendance at the meeting organized by
Professor Lesser Blum at the University of Puerto Rico in March 1996 in honor
of Bernard Jancovici.
The interest of Professors George Stell and Harold Friedman is appreciated.
The support of the National Science Foundation (through Grants CHE 93-11729 and
CHE 96-14495) has been essential.

\end{multicols}


\begin{references}
\bibitem{msalowt} See Eqs. (9.8) and (9.9) of Ref. \cite{lf} but note the
following corrections:
(a) In Eq. (9.8) the coefficient $-6 \sqrt{2}$ in the correction factor should
be: $-18$;
(b) In Eq. (9.9) the right-hand side should read:
 $(e^3/81 \pi T^{*3} ) e^{-1/\ts}[1 - \frac{9}{2}\ts + \cdots]$.
\end{references}
\end{document}